\documentclass[twocolumn]{revtex4}[12pt]
\usepackage{graphicx}
\begin{document}
\title{Propagation of shock structures in a high density plasma}
\author{ O. Sharif$^{*}$}
\address{Department of General Educational Development, Daffodil
International University, Dhanmondi, Dhaka-1207, Bangladesh\\
Email$^{*}$: omarsharif.ged@diu.edu.bd}

\begin{abstract}
A theoretical investigation has been made to study the cylindrical
and spherical electron-acoustic shock waves (EASWs) in an
unmagnetized, collisionless degenerate quantum plasma system
containing two distinct groups of electrons (one inertial
non-relativistic cold electrons and other inertialess
ultra-relativistic hot electrons) and positively charged static
ions. By employing well known reductive perturbation method the
modified Burgers (mB) equation has been derived. It is seen that
only rarefactive shock waves can propagate in such a quantum
plasma system. The effects of degenerate plasma pressure and
number density of hot and cold electron fluids, nonplanar
geometry, and positively charged static ions are responsible to
modify the fundamental properties of EASWs. It is also observed
that the properties of planar mB shocks are quite different from
those of nonplanar mB shocks. The findings of the present
investigation should be useful in understanding the nonlinear
phenomena associated with nonplanar EAWs  in both space and laboratory plasmas.\\

\end{abstract}

\maketitle
\section{Introduction}


The apparent interest for quantum plasmas has been retained their
interest due to their existence both in laboratory and space
plasma environments
\cite{Haas2003,Shukla2008,Eliasson2006,Khan2007,Misra2007,Sah2009b,Hossain2013,a1,a2,a3,a4}.
Generally, electron-acoustic (EA) waves (EAWs) occur in a plasma
environments (white dwarfs, neutron star, etc.) consisting of two
distinct temperature electrons (as "hot" and "cold" electrons)
\cite{Watanabe1977,Yu1983,Tokar1984,Mace1990,Kourakis2004,Mamun2004}.
EAWs are nothing but high-frequency electrostatic mode, for which
inertia is provided by the cold electron motion, while the
restoring force comes from the hot electron thermal pressure. The
positively charged ions may be safely assumed to be stationary,
simply maintaining the quasineutrality condition of the plasma
system. In such acoustic mode, the frequency lies in the range
between the plasma frequency of the cold and hot electron fluids.
Watanabe and Taniuti \cite{Watanabe1977} have first shown the
existence of the electron-acoustic (EA) mode in a plasma of
two-temperature (cold and hot) electrons. Some past decades, EA
waves has received a great deal of renewed interest not only
because of the two distinct group of electron plasma is very
common in laboratory experiments
\cite{Derfler1969,Henry1972,Kadomtsev1971,Armstrong1979,Sheridan1991}
and in space
\cite{Dubouloz1993,Pottelette1999,Singh2001,Ang2007,Barnes2003,Fedele2002a,Fedele2002b}
but also because of the potential importance of the EA waves in
interpreting electrostatic component of the broadband
electrostatic noise (BEN) observed in the cusp of the terrestrial
magnetosphere \cite{Tokar1984}, in the geomagnetic tail
\cite{Schriver1989}, in white dwarfs and neutron stars
\cite{Shapiro1983}, etc.

Now a days, researchers of plasma community gives great attention
to study the nonlinear behavior of astrophysical compact objects
e.g. white dwarfs, neutron stars, etc. The plasma particle number
density for such compact objects is so high (in white dwarfs it
can be of the order of $10^{30}$ $cm^{-3}$, even more)
\cite{Hossen2014,Hossen2014c,Hossen2014d,Hossen2014e} that the
electron Fermi energy is comparable to the electron mass energy
and the electron speed is comparable to the speed of light in a
vacuum. Chandrasekhar
\cite{Chandrasekhar1931a,Chandrasekhar1931b} presented a general
expression for the relativistic ion and electron pressures in his
classical papers. The pressure for electron fluid can be given by
the following equation

\begin{eqnarray}
&&P_e=K_en_e^\alpha, \label{Ch1}
\end{eqnarray}
where $n_e$ is the electron number density and
\begin{eqnarray}
&&\alpha=\frac{5}{3};~~K_e=\frac{3}{5}\left(\frac{\pi}{3}\right)^{\frac{1}{3}}
\frac{\pi\hbar^2}{m}\simeq\frac{3}{5}\Lambda_c\hbar c, \label{Ch2}
\end{eqnarray}
for the non-relativistic limit (where $\Lambda_c=\pi
\hbar/mc=1.2\times 10^{-10}~cm$, and $\hbar$ is the Planck
constant divided by $2\pi$). And
\begin{eqnarray}
&&P_e=K_en_e^\gamma, \label{Ch3}
\end{eqnarray}
where
\begin{eqnarray}
&&\gamma=\frac{4}{3};~~
K_e=\frac{3}{4}\left(\frac{\pi^2}{9}\right)^{\frac{1}{3}} \hbar
c\simeq\frac{3}{4} \hbar c, \label{Ch5}
\end{eqnarray}
for the ultra-relativistic limit
\cite{Chandrasekhar1931a,Chandrasekhar1931b,Chandrasekhar1935,Chandrasekhar1939,Mamun2010a,Mamun2010b,Hossen2014b,Hossen2014m,Hossen2014n,Hossen2014o}.

A large number number of works on relativistic degenerate quantum
plasma have been accomplished considering different acoustic
waves in the recent years
\cite{Shah2015a,Shah2015b,Shah2015c,Shah2015d,Hossen2014i,Hossen2015c,Hossen2014k,Ema2015,H2014a,H2014b,H2014c,H2014d,H2014e,H2015a,H2016a,H2016b}.
Han \textit{et al.} \cite{Han2013} investigate the existence of
electron-acoustic shock waves and their interactions in a
non-Maxwellian plasma with q-nonextensive distributed electrons .
Later on, Han \textit{et al.} \cite{Han2014} theoretically
investigated the nonlinear electron-acoustic solitary and shock
waves in a dissipative, nonplanar space plasma with superthermal
hot electrons. Sahu and Tribeche \cite{Sahu2013} considered
electron acoustic shock waves (EASWs) in an unmagnetized plasma
whose constituents are cold electrons, immobile ions and
Boltzmann distributed hot electrons and studied the effects of
several parameters and ion kinematic viscosity on the basic
features of EA shock waves. By considering quantum plasma
El-Labany \textit{et al.} \cite{EL-Labany2013} investigated the
effects of Bohm potential on the head on collision between two
quantum electron-acoustic solitary waves using the extended
Poincaré-Lighthill-Kuo method. Mahmood and Masood
\cite{Mahmood2008} illustrated that an increase in quantum
diffraction parameter broadens the nonlinear structure. Recently,
Sah \cite{Sah2009} demonstrated that the width, the amplitude,
and the velocity of electron-acoustic double layers, in three
component dense quantum plasmas consisting of stationary
background ions and two electron populations: one "cold" and the
other "hot", are significantly affected by the ratio of
unperturbed cold to hot electron densities. Again, the effect of
static ions is very common in plasma physics literature
\cite{Pakzad2011,Tribeche2010,Amour2012}. To the best of our
knowledge, none of the authors did consider the combine effects
of nonplanar geometry, effects of relativistic limits (i.e., both
non-relativistic and ultra-relativistic) and degenerate plasma
pressure which can significantly modify the propagation of
solitary and shock waves.
\section{Governing Equations}

We consider a cylindrical and spherical EA waves in an
unmagnetized, collisionless plasma, which is composed of
non-relativistic inertial cold electrons, both non-relativistic
and ultra-relativistic degenerate hot electron fluids, and static
positive ions. Thus at equilibrium, we have
$n_{i0}=n_{c0}+n_{h0}$, where $n_{s0}$ is the equilibrium number
density of the species $s$ ($s=c,\,h,\, i$ for cold electrons,
hot electrons, positive ions, respectively). The nonlinear
dynamics of the electrostatic waves propagating in such a
degenerate quantum plasma system is governed by the following
normalized equations

\begin{eqnarray}
&&\frac{\partial n_s}{\partial t}
+\frac{1}{r^\nu}\frac{\partial}{\partial r} (r^\nu n_s u_s) = 0,
\label{be1}\\
&&\frac{\partial u_c}{\partial t} + u_c \frac{\partial u_c}
{\partial r} +\frac{\partial \phi} {\partial
r}+\frac{K_1}{n_i}\frac{\partial n_i^\alpha}{\partial
r}-\frac{\eta}{r^\nu}\frac{\partial}{\partial
r}(r^\nu\frac{\partial u_i}{\partial r})=0,
\label{be2}\\
&&n_h \frac{\partial \phi}{\partial r}-K_2\frac{\partial
n_h^{\gamma}}{\partial r}=0,
\label{be3}\\
&&\frac{1}{r^\nu}\frac{\partial}{\partial
r}(r^\nu\frac{\partial\phi}{\partial r})=-\rho,
\label{be4}\\
&&\rho = \mu-n_c-(\mu-1)n_h, \label{be5}
\end{eqnarray}

\noindent where $\nu=0$ for one dimensional planar geometry,
$\nu=1$ (2) for nonplanar cylindrical (spherical) geometry, $n_s$
(s=c, h, i) is the the plasma species number density normalized
by its equilibrium value $n_{s0}$, $u_s$ is the plasma fluid
speed normalized by $C_{c}=(m_hc^2/m_c)^{1/2}$ with $m_h$ ($m_c$)
being the hot electron (cold electron) rest mass, $c$ is the
speed of light in vacuum, $\phi$ is the electrostatic wave
potential normalized by $m_hc^2/e$. Here $\mu\, (=n_{i0}/n_{c0})$
is the ratio of ion-to-cold electron number density. The time
variable ($t$) is normalized by ${\omega_{pi}}=\left(4 \pi
n_{c0}e^2/m_c\right)^{1/2}$, and the space variable ($x$) is
normalized by $\lambda_{s}=\left(m_hc^2/4 \pi
n_{c0}e^2\right)^{1/2}$. The coefficient of viscosity $\eta$ is a
normalized quantity given by
${\omega_{i}}\lambda_{mc}^2m_sn_{s0}$. We have defined
$K_1=n_{c0}^{\alpha-1}K_i/{m_e}{c}^2$ and
$K_2=n_{h0}^{\gamma-1}K_e/{m_e}{c}^2$.

\section{Derivation of modified Burgers Equation}

We derive a dynamical modified Burgers (mB) equation for the
nonlinear propagation of the EA waves by using equations
(\ref{be1})-(\ref{be5}). To do so, we employ a reductive
perturbation technique to examine electrostatic perturbations
propagating in the relativistic degenerate dense plasma system due
to the effect of dissipation, we first introduce the stretched
coordinates \cite{Maxon1974}:
\begin{eqnarray}
&&\xi=-{\epsilon}(r + V_pt),\,\,\,\,\,\tau={\epsilon}^{2}t,
\label{5b}
\end{eqnarray}

\noindent where $V_p$ is the wave phase speed ($\omega/k$ with
$\omega$ being the angular frequency and $k$ being the wave
number), and $\epsilon$ is a smallness parameter measuring the
weakness of the dissipation ($0<\epsilon<1$). We expand the
parameters $n_c$, $n_h$, $u_c$, $\phi$, and $\rho$ in power
series of $\epsilon$ as:

\begin{eqnarray}
&&n_c=1+\epsilon n_c^{(1)}+\epsilon^{2}n_c^{(2)}+ \cdot \cdot
\cdot, \label{6a}\\
&&n_h=1+\epsilon n_h^{(1)}+\epsilon^{2}n_h^{(2)}+ \cdot \cdot
\cdot, \label{6b}\\
&&u_c=\epsilon u_c^{(1)}+\epsilon^{2}u_c^{(2)}+\cdot \cdot \cdot,
\label{6c}\\
&&\phi=\epsilon\phi^{(1)}+\epsilon^{2}\phi^{(2)}+\cdot \cdot
\cdot, \label{6d}\\
&&\rho=\epsilon\rho^{(1)}+\epsilon^{2}\rho^{(2)}+\cdot \cdot
\cdot, \label{6e}
\end{eqnarray}

Now, expressing equations (\ref{be1})-(\ref{be5}) (using equation
(\ref{5b}), in terms of $\xi$ and $\tau$, and substituting
equations (\ref{6a})-(\ref{6e}), one can easily develop different
sets of equations in various powers of $\epsilon$. To the lowest
order in $\epsilon$, we have:
$u_c^{(1)}=V_p\phi^{(1)}/(V_p^2-K_1^\prime)$,
$n_c^{(1)}=-\phi^{(1)}/(V_p^2-K_1^\prime)$,
$n_h^{(1)}=\phi^{(1)}/K_2^\prime$,
$V_p=\sqrt{\frac{K_2^\prime}{\mu-1}+K_1^\prime}$, where
$K_1^\prime=\alpha K_1$ and $K_2^\prime=\gamma K_2$. The relation
$V_p=\sqrt{\frac{K_2^\prime}{\mu-1}+K_1^\prime}$ represents the
dispersion relation as well as the phase speed for the EA type
electrostatic waves in the degenerate quantum plasma under
consideration.

\noindent To the next higher order in $\epsilon$, we obtain a sets
of equations
\begin{eqnarray}
&&\hspace*{-1mm}\frac{\partial n_c^{(1)}}{\partial \tau}
-V_{p}\frac{\partial n_c^{(2)}}{\partial \xi}
-\frac{\partial}{\partial
\xi}[u_c^{(2)}+n_c^{(1)}u_c^{(1)}]-\frac{\nu
u_c^{(1)}}{V_{p}\tau}=0,
\label{NL6}\\
&&\hspace*{-1mm}\frac{\partial u_c^{(1)}}{\partial \tau}
-V_p\frac{\partial u_c^{(2)}}{\partial \xi}
-u_c^{(1)}\frac{\partial u_c^{(1)}}{\partial \xi}+\frac{\partial
\phi^{(2)}}{\partial
\xi}\nonumber\\
&&-K_1^\prime\frac{\partial}{\partial
\xi}[n_c^{(2)}+\frac{(\alpha-2)}{2}{(n_c^{(1)})}^2]-\eta\frac{\partial^2
u_{c}^{(1)}}{\partial \xi^2} =0,
\label{NL7}\\
&&\hspace*{-1mm}\frac{\partial \phi^{(2)}}{\partial \xi}
-K_2^\prime \frac{\partial}{\partial
\xi}\left[n_h^{(2)}+\frac{(\gamma-2)}{2}{(n_h^{(1)})}^2\right]=0,
\label{NL8}\\
&&\hspace*{-1mm}n_c^{(2)}+(\mu-1)n_h^{(2)}=0, \label{NL9}
\end{eqnarray}

\begin{figure}[t!]
\centerline{\includegraphics[width=6.5cm]{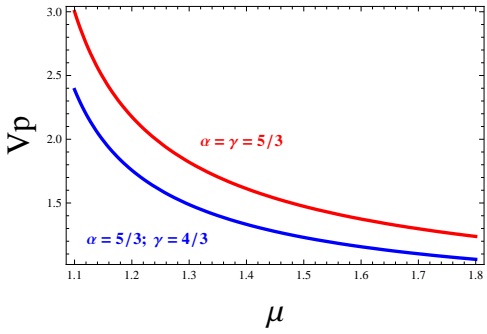}} \caption{(Color
online) Variation of phase speed $V_p$ with ion-to-cold electron
number density ratio $\mu$ for $u_0=0.1$.} \label{1}
\end{figure}

\noindent Now, combining equations (\ref{NL6})-(\ref{NL9}) we
deduce Burgers equation
\begin{eqnarray}
\frac{\partial\phi^{(1)}}{\partial \tau} + A \phi^{(1)}
\frac{\partial \phi^{(1)}}{\partial
\xi}+\frac{\nu\phi^{(1)}}{2\tau} = B \frac{\partial^2
\phi^{(1)}}{\partial \xi^2}, \label{EAB}
\end{eqnarray}
where
\begin{eqnarray}
&&\hspace*{-1mm}A=\frac{{(V_p^2-K_1^\prime)}^2}{2V_p}\left[\frac{(\gamma-2)(\mu-1)}{{K_2^\prime}^2}-\frac{3V_p^2+K_1^\prime(\alpha-2)}{{(V_p^2-K_1^\prime)}^3}\right],
\label{A}\\
&&\hspace*{-1mm}B=\frac{\eta}{2}. \label{B}
\end{eqnarray}
Here A and B are two constants and may be defined as nonlinearity
and dissipative constants respectively.
\begin{figure}[t!]
\centerline{\includegraphics[width=6.5cm]{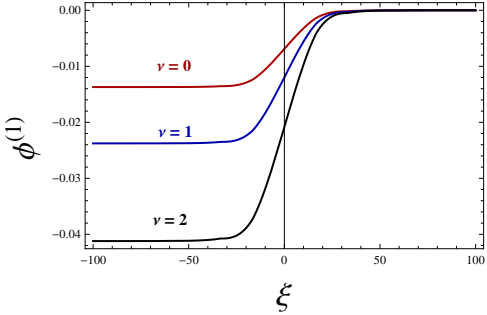}} \caption{(Color
online) Nonlinear shock waves are shown for different values of
$\nu$ when cold electron and hot electron fluids both are
nonrelativistic degenerate with parameters $u_0=0.01$ and
$\mu=0.93$} \label{3}
\end{figure}

\begin{figure}[t!]
\centerline{\includegraphics[width=6.5cm]{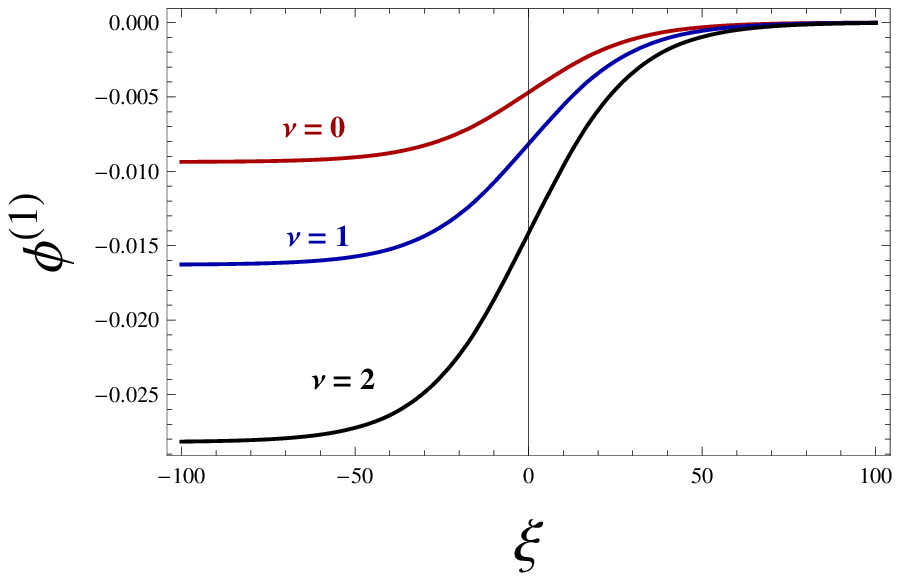}} \caption{(Color
online) Effects of cylindrical geometry on EA shock waves when
both cold electron and hot electron fluids are nonrelativistic
and degenerate ($\nu=1$, $u_0=0.01$ and $\mu=0.93$).} \label{4}
\end{figure}

\begin{figure}[t!]
\centerline{\includegraphics[width=6.5cm]{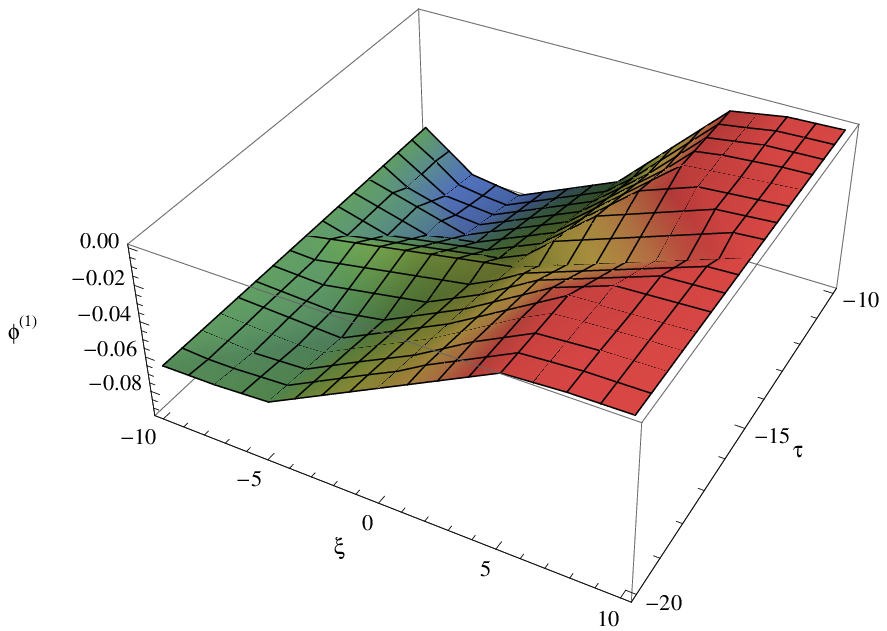}} \caption{(Color
online) Effects of cylindrical geometry on EA shock waves when
both cold electron and hot electron fluids are nonrelativistic and
degenerate ($\nu=1$, $u_0=0.01$ and $\mu=0.93$).} \label{5}
\end{figure}

\begin{figure}[t!]
\centerline{\includegraphics[width=6.5cm]{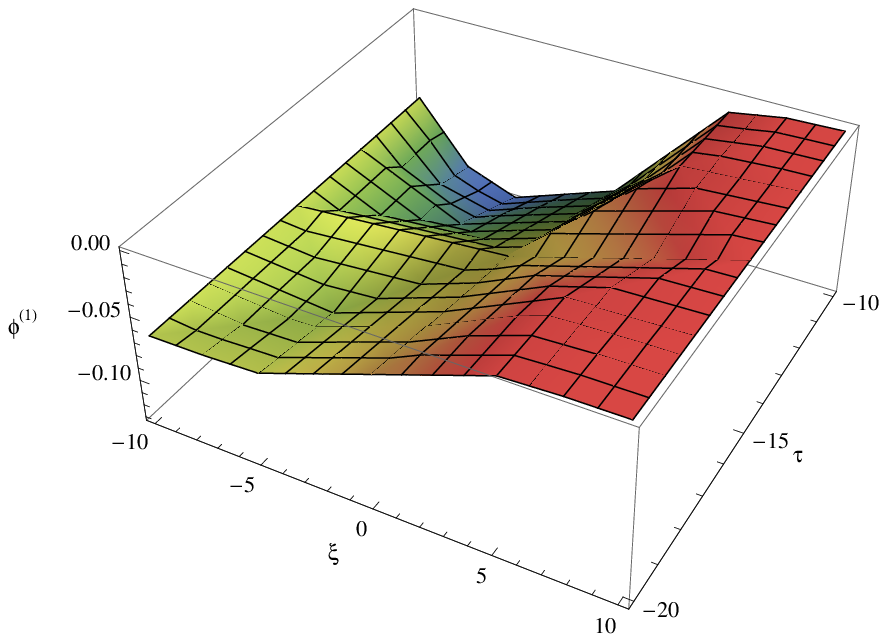}} \caption{(Color
online) Effects of spherical geometry on EA shock waves when both
cold electron and hot electron fluids are nonrelativistic and
degenerate ($\nu=2$, $u_0=0.01$ and $\mu=0.93$).} \label{5}
\end{figure}

\begin{figure}[t!]
\centerline{\includegraphics[width=6.5cm]{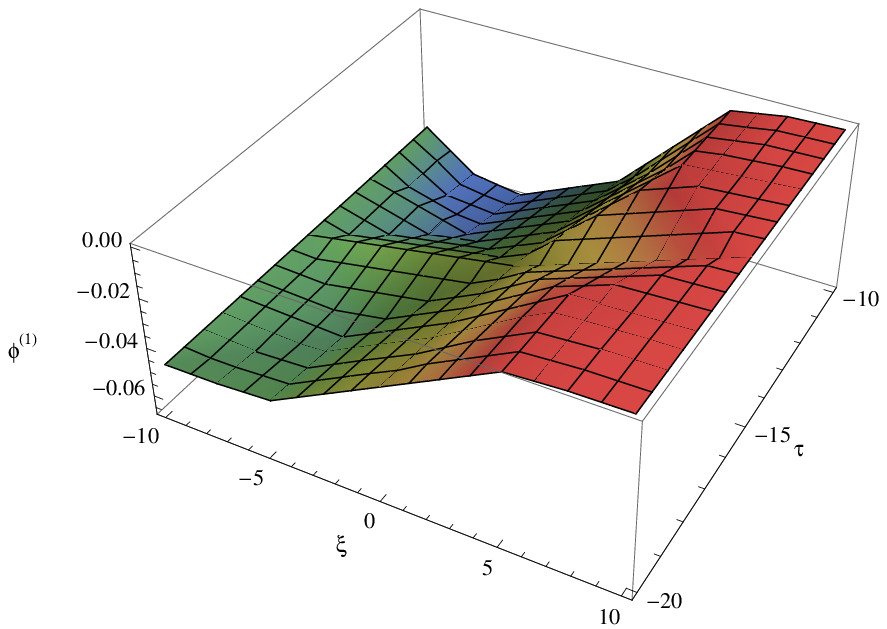}} \caption{(Color
online) Effects of cylindrical geometry on EA shock waves when
cold electrons being nonrelativistic degenerate and hot electrons
being ultrarelativistic degenerate ($\nu=1$, $u_0=0.01$ and
$\mu=0.93$).} \label{6}
\end{figure}

\begin{figure}[t!]
\centerline{\includegraphics[width=6.5cm]{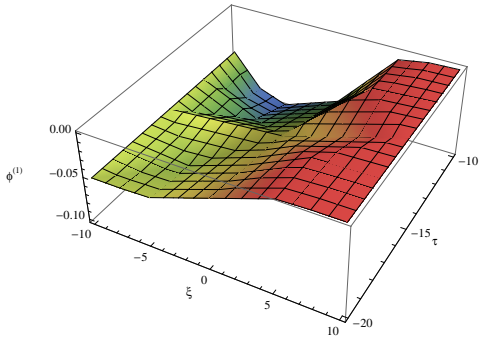}} \caption{(Color
online) Effects of spherical geometry on EA shock waves when cold
electrons being nonrelativistic degenerate and hot electrons
being ultrarelativistic degenerate ($\nu=2$, $u_0=0.01$ and
$\mu=0.93$).} \label{7}
\end{figure}


\section{Discussion and Results}

In this section, our first intention to numerically analyze the
Burgers equation. However, for clear understanding, we first
briefly discuss about the stationary shock wave solution for
equation (\ref{EAB}) with $\nu=0$, though the solution is similar
for both IA and EA waves (excluding the values of A and B). We
should note that for a large value of $\tau$, the term
$\frac{\nu\phi^{(1)}}{2\tau}$ is negligible. So, in our numerical
analysis, we start with a large value of $\tau$ (viz.
$\tau=-20$), and at this large (negative) value of $\tau$, we
choose the stationary shock wave solution of equation
(\ref{solK-dV}) [without the term $\frac{\nu\phi^{(1)}}{2\tau}$]
as our initial pulse. The stationary shock wave solution of this
standard Burgers equation is obtained by considering a frame
$\xi=\zeta-u_0\tau$ (moving with speed $u_{0}$ which is the ion
fluid speed at equilibrium) and the solution is
\cite{H2016m,H2016n,H2016o}

\begin{eqnarray}
{\rm
\phi^{(1)}}_{(\nu\rightarrow0)}=\phi_m^{(1)}\left[1-tanh\left(\frac{\xi}{\delta}\right)\right],
\label{solK-dV}
\end{eqnarray}
where $\phi_m^{(1)}=u_0/A$ and $\delta=2B/u_0$.

Now, we investigate the dynamical properties of EA shock waves in
terms of the intrinsic parameters of our model, namely the
ion-to-cold electron density ratio $\mu$ and the (cold electron)
kinematic viscosity $\eta$. It is important to not that the
dissipation term only depends on the electron kinematic viscosity
$\eta$. The shock profile is nothing but sudden increase or
decrease of the permanent profile of the waves
\cite{H2016p,H2016q,H2016r,H2016s}. It is observed that EA waves
are significantly modified when cold electron being
non-relativistic degenerate ($\alpha=\frac{5}{3}$) and hot
electron being ultra-relativistic degenerate
($\gamma=\frac{4}{3}$) than both cold and hot electron being
non-relativistic degenerate ($\alpha=\gamma=\frac{5}{3}$). It is
important to mention that modified Burgers equation derived here
is valid only for the limits $A\ne 0$, $A>0$ and $A<0$. It is
also important to note that for $\mu>\mu_c$ ($\mu_c=0.89$)
rarefactive shock waves are found but no compressive shock waves
exist at $\mu<\mu_c$. We have considered $u_{0}=0.01$ for our
numerical analysis of EASWs for the plasma system under
investigation here.

Figure 1 shows the variation of phase speed ($V_P$) with ion to
cold electron number density ratio $\mu$. It is found that the
phase speed decreases with the increasing values of $\mu$. It is
expected as the phase speed $V_p$ (derived from this considered
plasma) is higher for lower values of $\mu$ (see the expression
of $V_p$).  The variation of the rarefactive amplitude of shock
structures for both planar and nonplanar geometry is shown in Fig.
2 where cold electron and hot electron being nonrelativistic
degenerate. Figure 3 shows the variation of the rarefactive
amplitude of shock structures for both planar and nonplanar
geometry where cold electron being nonrelativistic degenerate and
hot electron being ultrarelativistic degenerate. The cylindrical
and spherical variation of the amplitude of EASWs for both
non-relativistic and ultra-relativistic limits is shown in Figs.
4-7. Finally, the results that we have found in this
investigation can be summarized as follows:

\begin{enumerate}

\item{The cylindrical and spherical plasma system under consideration supports only rarefactive shock waves
with negative potential, but no compressive shock waves exist.}

\item{The fundamental properties of EASWs are found to be significantly modified by the
relativistic parameters, nonplanar geometry and plasma particle
number densities.}

\item{It is observed that the phase speed (Vp) of these EA shocks inversely proportional to the square
root of ion to cold electron number densities ratio $\mu$.}

\item{It is also found that the phase speed ($V_p$) of EA waves decreases with the increasing values of $\mu$ (see Fig. \ref{1}).}

\item{It is observed that the amplitude of the shock is maximum for the spherical geometry, intermediate for cylindrical geometry, while it is minimum for the planar geometry (see Figs 2-3).}

\item{The amplitude of shocks proportional to the fluid speed $u_0$ but inversely proportional to the constant A.}

\item{From Figs. \ref{4}-\ref{7}, we observed that the amplitude of the nonplanar EA rarefactive shock waves is lower for
ultra-relativistic case than for non-relativistic case.}
\end{enumerate}

In conclusion, our simplified theoretical model represents a small
yet steady step towards the rigorous understanding of the
behavior of cylindrical and spherical EA shocks in degenerate
plasma environments, which appear to be of fundamental importance
in a wide range of astrophysical
\cite{Dubouloz1993,Pottelette1999,Singh2001,Ang2007,Barnes2003,Fedele2002a,Fedele2002b}
and laboratory scenarios
\cite{Derfler1969,Henry1972,Kadomtsev1971,Armstrong1979,Sheridan1991}.

\end{document}